# Network Analysis and Remote Application Control Software based on Client-Server Architecture

Ramya Mohan
University of Mumbai

## ABSTRACT
This paper outlines a comprehensive model to increase system efficiency, preserve network bandwidth, monitor incoming and outgoing packets, ensure the security of confidential files and reduce power wastage in an organization. This model illustrates the use and potential application of a Network Analysis Tool (NAT) in a multi-computer set-up of any scale. The model is designed to run in the background and not hamper any currently executing applications, while using minimum system resources. It was developed as open source software, using VB.Net, with a view to overcoming limitations of legacy systems and financial restrictions in small-to mid-level organizations like businesses and educational institutes. It is fully-customizable and serves as a simple and open-source alternative to existing software. The NAT relies on simple client-server architecture and uses remote access to monitor and maintain the computers on a network, for example logging off a user or shutting down a computer after a certain "idle" time, enabling and disabling applications, troubleshooting and so on. The NAT was tested in a laboratory and resultant data is presented, along with the results of a survey that was conducted among users.

## General Terms
Computer Networking.

## 1. INTRODUCTION
Computers have been used to automate mundane tasks and simplify complex ones. They have brought a wealth of information to our fingertips. Additionally, computers help analyze raw data and convert it into meaningful and reliable information. With over a billion users around the world, the computer has changed the way we communicate with one another.

Communication between two or more computers takes place due to a network. A "Computer Network'' is a collection of autonomous computers interconnected by a single technology. Connections can be made with the help of hardware like copper wires, fiber optic cables or waveforms like Bluetooth, infrared, WiFi and satellites. A computer network can be of any size- it could be a PAN (Personal Area Network) with a few computers connected, or a larger network like LAN (Local Area Network), MAN (Metropolitan Area Network) or Wide Area Network (WAN). Multiple networks can also be connected to form a bigger network- the internet is an example of such an extended or combined network. It is also possible to create networks in such a way that they allow access rights and restrictions. For example, the "Admin" of a network can have more control over a network than a standard employee. Virtual Private Networks or "VPNs" allow networks to connect to each other in a secure way. This is common in organizations where the office in one location can access data in the headquarters in another location.

The network settings and connection can be configured with the help of specialized software. [1]

One of the most commonly used technologies for network communication is the Transmission Control Protocol/Internet Protocol, also called TCP/IP (*Fig. 1*) [3] in short. It can be used for communication within a network, between multiple networks or the internet. It is used to define the related protocols and applications associated with itself, such as ARP, UDP, RARP, IPSec, HTTP, FTP, Telnet and many more. The TCP is a timed, connection-oriented, stateless layer that guarantees delivery of datagrams. At the sender's side, it deals with sorting the data into smaller packets, and the TCP layer at the receiver's side organizes these packets to form the original message. The IP layer provides important mechanisms like addressing and routing. The 32 bit address is used to identify the host machine and ensures that the packets are routed to their intended destination. The routes for packet forwarding and routing are determined by the IP Route Table. A list of local interfaces provides direct routes, and additional networks and gateways can be configured with indirect routing. The determination of direct routes is derived from the list of local interfaces. [1][2][3]

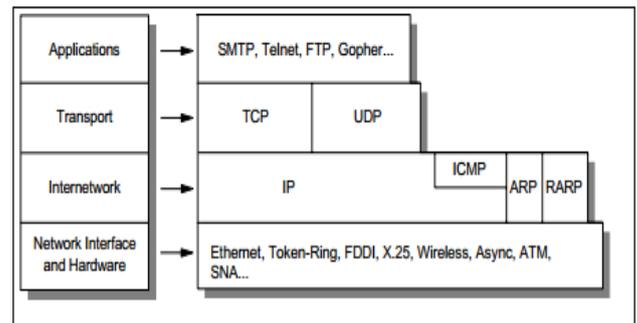

**Figure (1): TCP/IP Architecture**

## 2. EXISTING MODELS
With the massive flow of data across numerous networks in countless organizations, safeguarding sensitive information and automating tasks are important things to consider while setting up an infrastructure or network. Organizations are now investing massive amounts of resources to ensuring their network's security (protection against malware and viruses, data theft), network's efficiency (reducing wait times and providing a stable communication line), and making optimal use of bandwidth. A variety of personal as well as enterprise-level software are available for these purposes.

### 2.1.1 Wireshark®
Wireshark is one of the most popular open source softwares for analyzing a network. It is built using C and is an open source program that is constantly under development and review by experts across the world. Wireshark allows users to capture data from a network or analyze pre-recorded data that's stored in a file. Wireshark also allows creation of packet filters, along with setting preferences and saving them. Each action sequence is called a TCP session- it begins with a SYN (Synchronize Sequence Numbers) flag and ends with a FIN flag. (*Fig. 2*)[4]

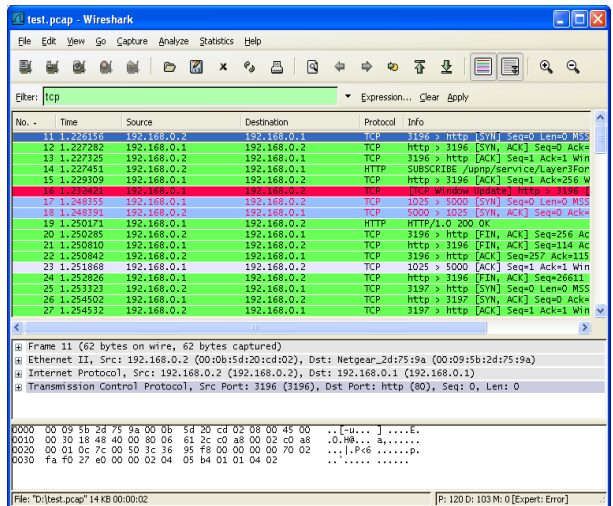

**Figure 2: Wireshark TCP Filter Display**

An acknowledgement flag is set on all the data packets that the client transmits after the SYN flag. This helps inspect incoming and outgoing packets, set privileges and restricts and can effectively function as a firewall. Wireshark is freely available under an open license, with a full set of features, for use in both personal and enterprise networks. [5]

### *2.1.2 Remote Desktop Protocol*

Remote Desktop Protocol®, developed by Microsoft, is software for client-server communication. It is based on the T.120 family of protocols and supports use of multiple channels along with creation of virtual communication networks. It projects the image of a remote desktop to a client, who can then operate it as his or her own desktop. On the client side, the packets received by the RDP and then translated into a Microsoft Graphics Device Interface. On the server side, the RDP utilizes its onscreen keyboard and mouse to interpret incoming client-side signals. In addition to facilitating communication, the RDP also provides encryption for keyboard and mouse. [6] It is free to use, supports all versions of Windows, along with Linux and Mac OSes. It uses 56-bit and 128 bit-encryption using RSA's RC-4 cipher, consumes low bandwidth, sound and printer redirection and load balancing. [7]

Not all such softwares are freeware. For large scale organizations, IT solutions are provided by other companies like 1E's NightWatchman®, Data Synergy's PowerMAN®, Verismic's Power Manager® and others. In addition to monitoring networks, they also assist with legal issues like software licenses and usage rights, energy efficiency, and task-automation. Such comprehensive packages also eliminate the technical expertise required by personnel or hassle involved when an organization has to personally deploy and manage these tools.

While full-fledged companies can have anywhere between 1000-10,000 computers, smaller ones often have a simple network of around 50-100 computers. Data transfers and bandwidth consumed is also significantly lesser, making enterprise plans an expensive option.

## 3. APPLICATION DESCRIPTION

The NAT is targeted for use in independent businesses, libraries, schools and colleges. It requires no prior networking skills or knowledge about technical terms, it consumes minimal memory space, has user-friendly GUI, is easy to set-up and is free to use. It makes use of simple client-server architecture. One file is installed on the "server", which would be the admin's computer and the other file is installed on every machine in the network. On installing the server side application, a list of all the active computers on the network is displayed and the admin can install the server side application on each of them. This then allows the server side application to communicate with the client side application. Information about incoming and outgoing packets, data consumption, idle time and active applications are displayed.

The NAT application has been developed using VB.Net, a robust programming language by Microsoft. It is a type-safe language, i.e., it can access only the memory locations it is explicitly granted permission to access and supports inheritance, multithreading and parameterized constructors. It also uses the framework's support of Namespaces to access libraries for the application. Some of the Namespaces used include: [8] [9]

1. **System:** Includes essential classes and base classes for commonly used data types, events, exceptions and so on.

2. **System.Data:** Includes classes which let us handle data from data sources.

3. **System.Data.OleDb:** Includes classes that support the OLEDB .NET provider.

4. **System.Diagnostics:** Includes classes that allow to debug the application and to step through the code.

Another benefit is tweaking softwares. Many applications run in the background, increasing CPU and memory usage and thus increasing power consumption. The admin can end unnecessary processes, install applications on multiple computers and provide remote troubleshooting assistance. The admin can also remotely install security measures, run scans, install software updates and patches and system upgrades at a time when it is not intruding on work-related tasks and thus affecting productivity. On a busy network, or one over which many users plug-in peripherals like CDs and USBs or download files, there is a high risk of malware. This could access sensitive documents or personal data, slow down the machine, infect the network or over clock the RAM. When this happens, the NAT receives a notification of an attack and isolates the infected computers to prevent the virus from spreading across the network. At this point, the admin can run scans and eliminate the virus by keeping the system in a sandbox and reconnect it to the network once the threat has been removed.

It's not uncommon to leave the computers running when they are not in use. Over a period of time, this leads to a significant amount in electricity bills or reduced computer performance simply due to negligence. With this NAT, the admin can review the idle time (keyboard or mouse movement) and send the computer into hibernation mode. After work hours, they can set them to shut down or shut down after finishing software updates, scans, and other maintenance tasks. [10] All the information collected by the NAT is stored in a database which is managed using MySQL. This allows the admin to review computer performance, efficiency, energy usage, data transfers and log reports of viruses, crashes and other data which help in determining the overall effectiveness of the software and functioning of the network in the organization.

## 4. SYSTEM ARCHITECTURE:

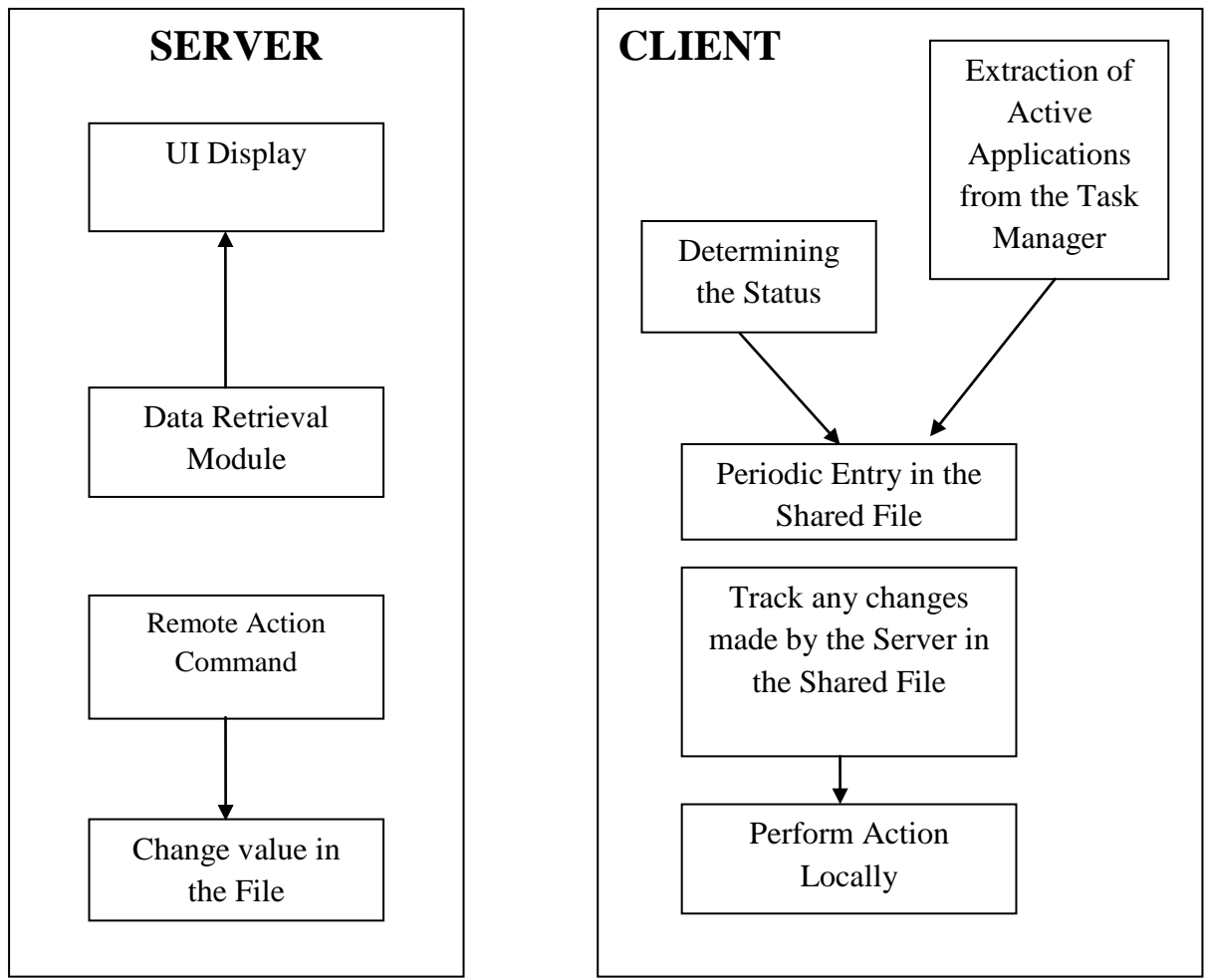

**Figure 3: System Architecture**

## 5. WORKING OF THE SYSTEM
The system relies on 2 components: The Server (Admin's computer) and the Client (computers to be monitored).

### 5.1 Server-side Functionality
1. **Data Retrieval Module:** Deals with definition of the OLEDB communication and the retrieval of the client-side data in the shared file. It makes use of the SQL statements to extract the needed information and sends it to the Display module.

2. **Display:** It organizes the available data into tables or lists.

3. **Remote Action:** This provides the Action buttons that initiate Shutdown or Restart in the computer the Admin selects from the shared file. It also gets the corresponding IP address of the computer selected and provides it to the next module.

4. **Change Value in the File:** This eliminates the need for the Microsoft .NET framework to be available on the client machines. The Server simply makes a specific entry in the shared file and this prompts the Client to invoke its own Shutdown.exe.

### 5.2 Client-side Functionality
1. **Extraction of Active Applications:** This module extracts the active applications from the Task Manager and enters it into the database.

2. **Detect Idle Time:** Monitors keyboard and mouse inactivity. [11]

3. **Determining the Status:** Depending upon the data entered by the above module, the status of the PC is determined as BUSY/IDLE. This keeps a timer for idle state, and the value (20 minutes, 30 minutes, an hour) can be set by the Admin. If the count is 0, the status is IDLE, else BUSY.

4. **Periodic Entry in the Shared File:** Helps track the state of the computer and its application dynamically and invokes the modules referenced above.

## 6. DATA FLOW DIAGRAMS:

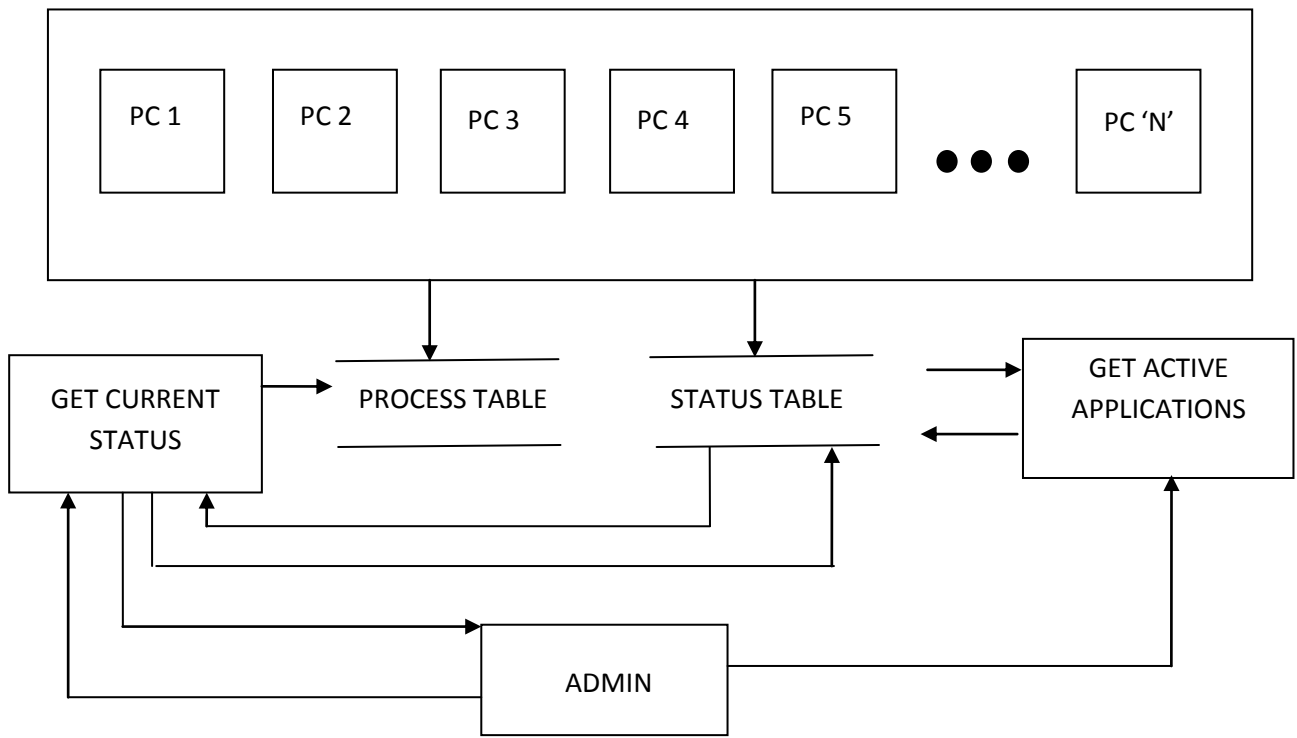

**Figure 3: Check PC Status and retrieve active applications**

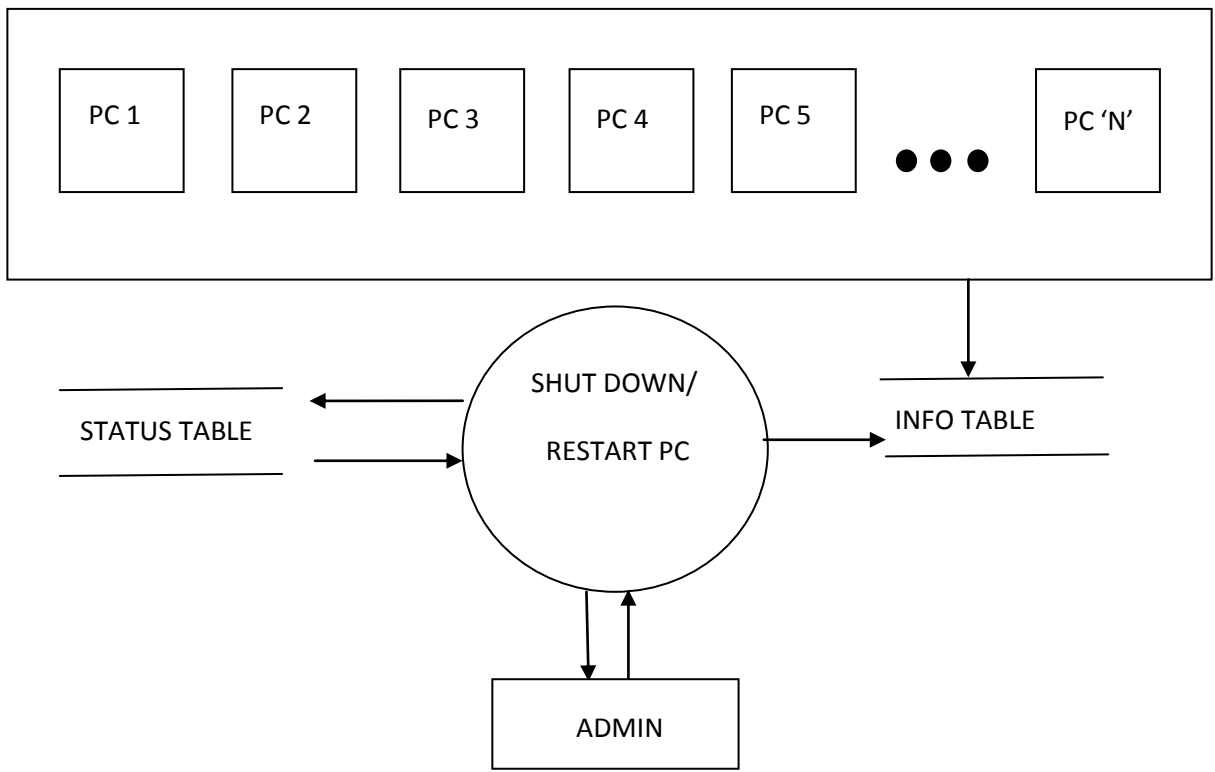

**Figure 4: DFD: Shut down and Restart Computer**

## 7. SEQUENCE OF EVENTS
The sequences of events of the NAT are shown below:

1. System Startup. Retrieval of information from Shared File.
2. Network Scanning.
3. Displaying the corresponding list to the administrator.
4. Selection of a particular workstation and accessing its data traffic, active applications etc.
5. Depending upon the status and other information, selecting an action (if needed).
6. System sending request to perform the specified action.
7. Response/Acknowledgement to the Administrator Request: The Client switching OFF/RESTARING it serves as an acknowledgement.
8. Further Notification to the Admin: The updated list does not contain an entry of that PC which indicates it has been switched OFF/RESTARTED.

## 8. IMPLEMENTATION
The first implementation of this software was in the engineering department of the host college. The department had 6 labs, each having between 30-50 computers. The NAT was tested on computers in two labs and on a total of 110 computers, of which 47 were CRT displays and the rest were LCDs. The computer configurations of the CRT displays were:

1. Display: Compaq CRT screens.
2. Processor: Intel® Pentium® 4, CPU 2.8 Ghz.
3. Memory: 40 GB SATA hard disks; 512 MB Ram
4. OS: Windows XP, SP 1.

The configurations of the LCDs were:

1. Display: Samsung LCD screens.
2. Processor: Intel®Pentium® D, CPU 3.0 GHz.
3. Memory: 80 GB hard disks, 1 GB RAM
4. OS: Windows 7 Ultimate, SP 1.

The average power consumption of the CRT computers was 210 watts and the average power consumption of the LCD computers was 160 watts. 38 out of 180 computers had at least 1 suspicious application active. 14 of the computers had malware, outdated anti-virus softwares, and memory-hungry processes. It was also observed that at any given point, at least 10 computers had only their displays turned off, but were still active. The server side of the MS application was first installed on the admin's computer and authorized by the server. Then the client side of the application was installed on the rest. We then performed the following operations:

1. Installed programming tools on the machines,
2. Ended unnecessary active applications,
3. Uninstalled programs,
4. Set browsing restrictions,
5. Ran a system scan remotely,
6. Logged off systems with a 10 minute period of no keyboard or mouse activity.

## 9. REQUIREMENTS
The minimum requirements for the NAT are given below:

1. Operating System: Windows 2000, Windows XP, Windows Vista, Windows 7.
2. Visual Studio 2005 or 2008, with SQL Server Express enabled (for coding). Server Side only.
3. Database Access. Server Side only.
4. Minimum RAM of 512 MB on Server Side, 256 MB for the Client Side.
5. A typical LAN network.

## 10. RESULTS
After testing the software for 7 working days, a survey was conducted a survey of 160 individuals who used the computer systems, in order to gauge the effectiveness of the NAT and the efficiency of the computers. Two admins and 2 professors of the computer department were also included in the survey. The questionnaire was prepared to gain insight into a user's experience with the application. The users were given a brief description of the NAT and asked to evaluate their experience on a scale of 1-10. Questions asked only to the admin and professors are marked by an asterisk (*). The percentages calculated from the feedback are as shown below:

**Table 1. Survey results**

| Criteria | User Satisfaction |
|---|---|
| Computer Speed | 79% |
| Malware Detection | 83% |
| Task Automation* | 75% |
| Accuracy | 100% |
| Energy Efficiency* | 61% |
| User Interface* | 82% |
| Overall Satisfaction | 92% |

## 11. FUTURE SCOPE
With the increasing dependency on computers in our day to day lives and the growth of networks as a path of communication, network analysis tools are essential to ensure safe and secure data transmission.

Such NATs can be implemented in schools, workspaces, homes and companies. Additional functionalities can be implemented to enhance the software's uses. For example, the software could be developed to provide:

1. Data encryption and data decryption,
2. Support plug-ins to analyze certain processes or machines,
3. Improve database connectivity and repositories to check for new viruses and potential threats,
4. More options like flags, application privileges, tools to increase flexibility.

The NAT can be made platform independent as well, since its functionality is limited to only machines running Windows OS.